\documentclass[acus]{JAC2003}

\usepackage{bm}
\usepackage{multicol}
\usepackage{graphicx}
\usepackage{latexsym}
\usepackage{amsmath}
\usepackage[draft]{hyperref}

\hypersetup{linktocpage=false,
            pdfborder={0 0 0},
            colorlinks=false,
            unicode=true}

\usepackage{txfonts}
\usepackage{cancel}
\usepackage{bbding}
\usepackage{subfig}
\usepackage{float}
\usepackage{subfloat}
\usepackage{caption}
\usepackage[section]{placeins}
\DeclareGraphicsExtensions{.pdf,.png,.jpg}

\usepackage[hang]{footmisc}
\begin{document}

\title{PROTON ACCELERATION BY A RELATIVISTIC LASER FREQUENCY-CHIRP DRIVEN PLASMA SNOWPLOW
	\thanks{ Work supported by the National Science Foundation under NSF-PHY-0936278}\\[-.8\baselineskip]}

\author{A. A. Sahai, \thanks{aakash.sahai@duke.edu}, T. C. Katsouleas, ECE, Duke University, Durham, NC, 27708 USA,\\
        		Robert Bingham, RAL, Didcot, OX11 0QX, UK,\\
		Frank S. Tsung, Adam Tableman, Michail Tzoufras, Warren B. Mori,\\ 
		Department of Physics and Astronomy, UCLA, Los Angeles, CA, 90095 USA}

\maketitle

\begin{abstract}
We analyze the use of a relativistic laser pulse with a controlled frequency chirp incident on a rising plasma density gradient to drive an acceleration structure for proton and light-ion acceleration. The Chirp Induced Transparency Acceleration (ChITA) scheme is described with an analytical model of the velocity of the snowplow at critical density on a pre-formed rising plasma density gradient that is driven by a positive-chirp in the frequency of a relativistic laser pulse. The velocity of the ChITA-snowplow is shown to depend upon rate of rise of the frequency of the relativistic laser pulse represented by $\frac{\epsilon_0}{\theta}$ where, $\epsilon_0 = \frac{\Delta\omega_0}{\omega_0}$ and chirping spatial scale-length, $\theta$, the normalized magnetic vector potential of the laser pulse $a_0$ and the plasma density gradient scale-length, $\alpha$.
We observe using 1-D OSIRIS simulations the formation and forward propagation of  ChITA-snowplow, being continuously pushed by the chirping laser at a velocity in accordance with the analytical results. The trace protons reflect off of this propagating snowplow structure and accelerate mono-energetically.
The control over ChITA-snowplow velocity allows the tuning of accelerated proton energies.
\end{abstract}

\section{Introduction}
Chirp-pulse amplification (CPA) technique has allowed construction of compact very high-intensity femtosecond laser pulses enabling experimental observation of physics of relativistic Laser-Plasma interactions. The knowledge of these processes of laser-plasma interactions is being applied to the development of compact proton and ion accelerator systems. These compact and relatively cheaper accelerators may allow larger adoption of hadron-therapy for cancer treatment by the medical community. In addition to medical applications these ultra-short and ultra-intense proton beams may be potentially applied to Fast-ignition fusion research, particle production and compact alternatives to conventional particle injectors. Amongst the various desirable characteristics of an accelerated beam, mono-energeticity and collimation are crucial. However, the existing laser-plasma acceleration techniques do not demonstrate the capabilities to meet the requirements for many of these applications. The scaling laws with picosecond and femtosecond laser pulses and different plasma / target parameters have been extensively studied\cite{scaling-laws-2006}\cite{scaling-laws-2007}.

\begin{figure}
	\begin{center}
   	\includegraphics[width=3.25in]{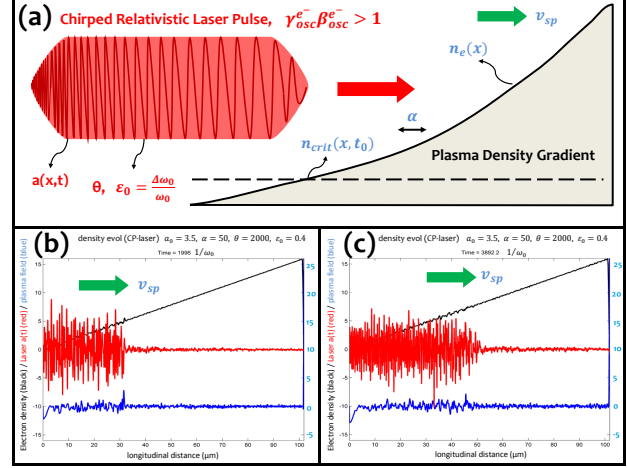}
	\end{center}
\caption{(a) Conceptual diagram depicting a chirped laser pulse propagating towards a rising plasma density gradient. (b) 1-D OSIRIS\cite{osiris-code-2002} simulations showing ChITA-snowplow in electron density and the corresponding electric field at $t=1996\frac{1}{\omega_0}$ with a circularly-polarized (CP) laser of normalized magnetic vector potential, $a_0=3.54$, laser frequency chirp parameters of $\epsilon_0=0.4$, $\theta=2000\frac{c}{\omega_0}$ ($\Rightarrow 850fs$) and plasma density gradient scale length of $\alpha=50\frac{c}{\omega_0}$ ($\simeq 6.4\mu m$). (c) The ChITA-snowplow at a later time, $t=3892.2\frac{1}{\omega_0}$, propagating further longitudinally into the rising plasma density gradient. The longitudinal distance is represented in $\mu m$ assuming a Ti:Sapphire laser with a wavelength of $0.8\mu m$}
\label{Snowplow-concept}
\end{figure}

The existing regimes include Target Normal Sheath Acceleration (TNSA), Collision-less Electrostatic Shock acceleration (CESA) and Radiation Pressure Acceleration (RPA). The CESA and RPA techniques offer potential to generate relatively mono-energetic proton beams. However, the recent experiments using the CESA and RPA techniques have had to rely on high-energy $CO_2$ lasers (picosecond pulse lengths) which are gas (active medium) laser and unlike the Ti:Sapphire crystal CPA lasers, may not be conducive to miniaturization.

In this paper, we present preliminary analysis of a unique scheme that accelerates the protons and light ions mono-energetically by reflecting them off of a continuously driven acceleration structure which we refer to as a ChITA-{\it Snowplow}. The controlled positive frequency chirp of a relativistic laser pulse interacting with the critical density layer of the plasma {\it continuously} drives the snowplow longitudinally forward. The control of frequency chirp is assumed to be experimentally realizable considering the fact that CPA laser pulses have residual frequency chirp. And, there is also a significant research dedicated to higher-harmonic generation. The radiation pressure of the laser pushes the critical layer electrons longitudinally forward, displacing them from the heavier background ions, creating a region of charge imbalance which develops the snowplow electric field. The propagating snowplow potential (when above a threshold) reflects the protons and light ions ahead of it to twice its velocity. The laser-plasma parameters of this scheme can be used to control the snowplow velocity and thereby tune the proton energies. 

\section{ChITA scheme - analytical model}

We analyse the propagation of a frequency-chirped relativistic laser pulse into a rising plasma density gradient. The laser is modeled with normalized magnetic vector potenital $a_0$ ($=\frac{e|\vec{A}|}{m_ec^2}=\frac{\vec{p}_\perp^{~e}(t)}{m_ec}$) and frequency of $\omega_0$ at its head. The rate of rise of the frequency of the laser pulse is controlled and can be described by frequency-rise chirp-fraction $\epsilon_0$ (eq.\ref{laser-chirp}) and its rise scale-length $\theta$. For every time interval of $\frac{\theta}{c}$ the laser frequency increases by $\Delta\omega_0$.

\begin{align}
\nonumber \epsilon_0 &= \frac{\Delta\omega_0}{\omega_0} \\
\newline \epsilon(x,t) &= \epsilon_0\left(\frac{ct-x}{\theta}\right)H(ct-x)
\label{laser-chirp}
\end{align}
\noindent The Heaviside step function $H(ct-x)$ ensures that the frequency chirping effect is observed at a point x in space only after the head of the pulse has reached that point, x (when $(ct-x)>0$).

The dispersion relation for the {\it Transverse mode} of electromagnetic radiation (pure sinusoid of frequency $\omega_0$) interacting with collision-less cold-plasma of electron-plasma frequency, $\omega_p = \sqrt{\frac{4\pi e^2 n_e}{m_e}}$, is in eq.\ref{transverse-mode-dispersion}. When the electromagnetic radiation encounters the condition where $\omega_0=\omega_p$, its wave-vector, $k=0$ and it cannot propagate into plasma beyond the critical density, $n_c=\frac{m_e\omega_0^2}{4\pi e^2}$.
\begin{equation}
\omega_0^2=\omega_p^2+k^2c^2
\label{transverse-mode-dispersion}
\end{equation}

However, if the intensity of the laser field is high enough such that it can cause the plasma electrons to undergo relativistic quiver oscillations, increasing their mass by the Lorentz factor, laser pulse can propagate beyond the critical density. This is called Relativistically Induced Transparency\cite{rit-prl-1971}. The Lorentz factor of the electrons quivering relativistically in the laser field is, $\gamma_{\perp}^{e^-}=\sqrt{1+\frac{\vec{p}_{\perp}.\vec{p}_{\perp}}{m_e^2c^2}}=\sqrt{1+a_0^2}$.
This relativistic electron mass increase thereby changes the plasma frequency to $\omega_p^{\gamma} = \sqrt{\frac{4\pi e^2 n_e}{m_e \sqrt{1+a_0^2}}}$.

In the ChITA scheme frequency of the laser pulse is increased to enable the laser pulse to propagate further into the rising plasma density, with increasing $n_e(x)$ resulting in the increase in electron fluid oscillation frequency, $\omega_{pe}(x)$. If the laser frequency $\omega(x,t)=\omega_0(1+\epsilon(x,t))$ is increased such that the laser field can only be shielded by plasma electron fluid oscillating at a higher $\omega_{pe}$, then the laser with increasing frequency can propagate further longitudinally. We refer to this process as {\it Chirp Induced Transparency}. The control of the frequency rise is important to maintain resonance with plasma electron fluid, to enable optimal transfer of energy from laser to plasma electrons. The propagating critical layer which we refer to as snowplow is depleted of electrons which being pushed by the laser radiation-pressure pile-up in a steepened density just beyond the critical layer, giving rise to the snowplow electric-field. The propagating snowplow potential reflects and accelerates the protons and light ions ahead of it. We refer to this acceleration method as {\it Chirp Induced Transparency Acceleration} (ChITA). 

The laser pulse incident at the plasma interface is assumed to be a long flat pulse with approximately 5 laser periods ($30\frac{1}{\omega_0}$) in the rise and fall of the pulse. The laser pre-pulse (picoseconds to nanosecond long) creates the heavy-ion metal plasma by ablation. The diffusion of plasma away from metal-air interface over the pre-pulse time-scales creates the plasma density gradient, before the main pulse arrives.

The plasma density profile created by laser pre-pulse ablating the metallic foil (``over-dense" plasma) is assumed to be linearly rising with the rise-scale length of $\alpha$ as in eq.\ref{plasma-density-gradient}.
\begin{equation}
n_e(x) = n_c\left(\frac{x}{\alpha}\right)
\label{plasma-density-gradient}
\end{equation}
\noindent This model simplifies the analysis and understanding of the interaction process. The scale length can be experimentally controlled by varying the intensity and duration of the pre-pulse.  

In consideration of the relativistic laser frequency-chirp (time varying frequency, $\omega(t)$) and its interaction with the rising plasma-density gradient, the plasma frequency at the relativistically corrected critical density ($k=0$) is a function of space and time, as in eq.\ref{plasma-frequency-variation}. 
\begin{align}
\omega(x,t) = \omega_p^{\gamma} &= \sqrt{\frac{4\pi e^2 n_e(x)}{\gamma m_e}} = \sqrt{\frac{4\pi e^2 n_c(x/\alpha)}{m_e \sqrt{1+a_0^2}}} \\ 
\newline \omega(x,t) = \omega_0\left(1+\epsilon(x,t)\right) &= \sqrt{\frac{4\pi e^2 n_c}{m_e}}\sqrt{\left(\frac{(x/\alpha)}{\gamma}\right)}\\
\newline \left(1+\epsilon_0\left(\frac{ct-x}{\theta}\right)\right)^2 &= \frac{(x/\alpha)}{\gamma}
\label{plasma-frequency-variation}
\end{align}

By solving eq.\ref{plasma-frequency-variation}, we obtain the time-dependent expression for the location of the moving critical density driven by the frequency-chirp. This is the density at which the laser is shielded and stopped by the plasma electrons and thereby at this density the energy exchange between the laser and plasma is through a resonant process. The electron quiver frequency in the laser electric field is equal to the plasma electron fluid frequency.  
\begin{align}
\nonumber & x_{sp}(t) =\\ 
\newline & \frac{2 c t \alpha  \gamma  \epsilon_0 ^2 + 2 \alpha  \gamma  \epsilon_0  \theta + \theta ^2 \pm \theta  \sqrt{4 c t \alpha  \gamma  \epsilon_0 ^2 + 4 \alpha  \gamma  \epsilon_0  \theta + \theta ^2}} {2 \alpha  \gamma  \epsilon_0 ^2}
\label{x-sp-location}
\end{align}

Partially differentiating eq.\ref{x-sp-location} with respect to time, we obtain the velocity of the chirp driven snow-plow, eq.\ref{v-sp-velocity}.
\begin{align}
& v_{sp}(t)=c-\frac{c \theta }{\sqrt{4 c t \alpha  \gamma  \epsilon_0 ^2+\theta  (4 \alpha  \gamma  \epsilon_0 + \theta )}}
\label{v-sp-velocity}
\end{align}

\noindent We substitute the parameters of Fig.\ref{Snowplow-concept}, into the ChiITA-snowplow velocity expression, eq.\ref{v-sp-velocity}. This gives us a ChiITA-snowplow velocity at the time $t=0$, of $v_{sp}=0.066c$.

\section{Simulation results}
To verify the formation of ChITA-snowplow and its frequency-chirp driven propagation we use OSIRIS Particle-In-Cell (PIC) 1D code for simulating a chirped laser pulse interacting with plasma density gradient. The background plasma ions are assumed to be fixed. The simulation is Eulerian and setup with 20 cells per $\frac{c}{\omega_0}$, 40 particles per cell and with the normalization of $\frac{\omega_p}{\omega_0}=1$. The results presented in this paper are with a circularly polarized (CP) laser pulse. Linearly polarized (LP) laser pulse has also been verified. The simulation evolves with time-steps of $\Delta t=0.0499\frac{1}{\omega_0}$ and a sliding window time average over
one laser period is applied before fields and phase-space data are dumped. The trace protons are modeled using test species at $10^{-4}\times n_c$.

\begin{figure}
	\begin{center}
   	\includegraphics[width=3.25in]{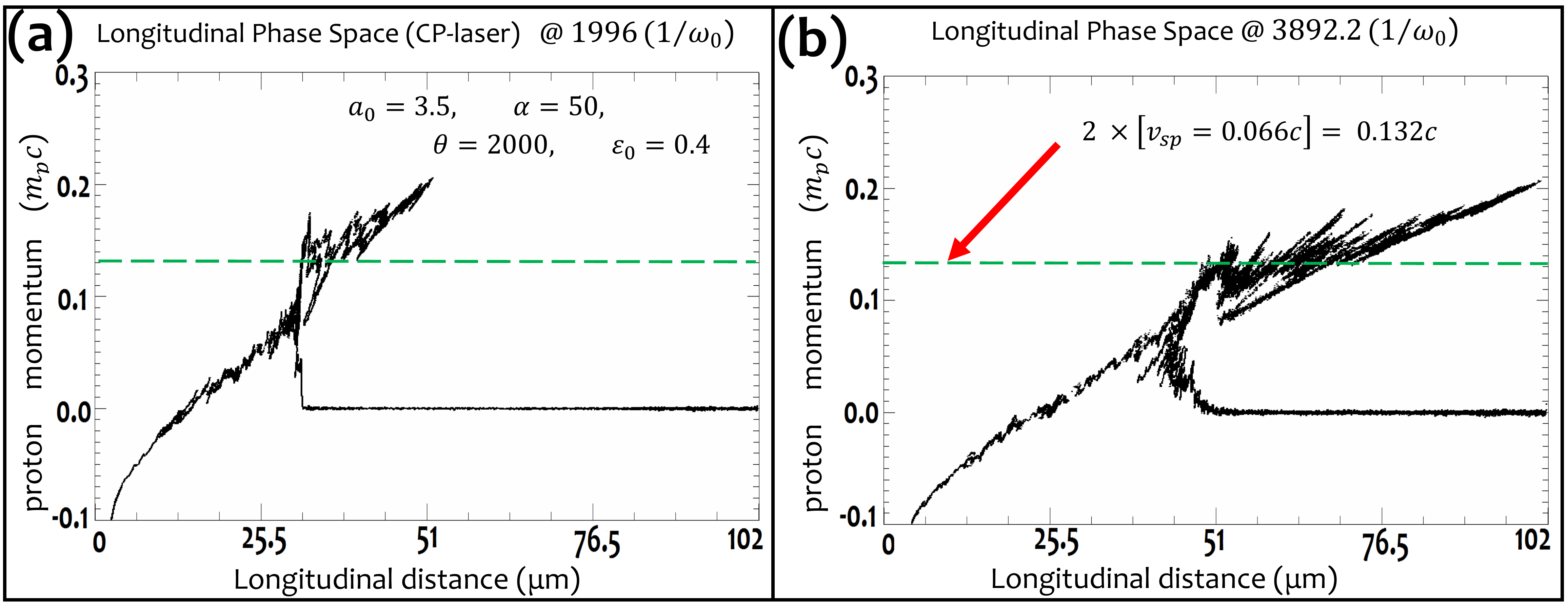}
	\end{center}
\caption{(a) Longitudinal phase space of the trace protons (setup as test particles at $10^{-4}\times n_c$) being reflected by the ChITA-snowplow potential corresponding to the density snapshot, Fig.\ref{Snowplow-concept}(b), with identical laser-plasma parameters at $t=1996\frac{1}{\omega_0}$. An initial bunch is launched at momentum of $0.19m_pc$ by the rise-time effects \cite{ieee-pac-2011} but when the pulse rises to its flat part ChITA-snowplow drives and protons are launched at momentum of $0.132m_pc$. (b) $t=3892.2\frac{1}{\omega_0}$,Fig.\ref{Snowplow-concept}(c). ChITA-snowplow has advanced longitudinally and continues to reflects protons mono-energetically at a momentum of $0.132m_pc$.}
\label{Chirp-Snowplow-1D-long-phase-space}
\end{figure}

Using 1-D simulations we find that the ChITA snowplow for the laser-plasma conditions propagates approximately at the analytically predicted velocity, $v_{sp}=0.066c$. It can also be observed from the trace-proton longitudinal phase space Fig.\ref{Chirp-Snowplow-1D-long-phase-space}(a)(b), that the protons are reflected off the propagating snowplow at $2\times v_{sp}=0.132c$, thereby gaining a momentum of $0.132m_pc$. In Fig.\ref{Chirp-Snowplow-1D-long-phase-space}(a), an initial bunch which is launched close to $0.19m_pc$, due to laser-pulse rise-time effects can also be observed \cite{ieee-pac-2011}. In the simulations when the laser is not chirped, $\epsilon_0=0$, the propagation occurs initially only during the time-period that can be attributed to rise-time snowplow \cite{ieee-pac-2011}. There is negligible forward propagation of the critical layer in the flat-part of the laser pulse.

\end{document}